\documentclass[conference]{IEEEtran}
\IEEEoverridecommandlockouts
\usepackage{cite}
\usepackage{amsmath,amssymb,amsfonts}
\usepackage{algorithm}
\usepackage{algorithmic}
\usepackage{graphicx}
\usepackage{textcomp}
\usepackage{booktabs}
\usepackage{multirow}
\usepackage{tikz}
\usetikzlibrary{positioning,arrows.meta}
\usepackage{xcolor}
\usepackage{url}
\def\BibTeX{{\rm B\kern-.05em{\sc i\kern-.025em b}\kern-.08em
    T\kern-.1667em\lower.7ex\hbox{E}\kern-.125emX}}
\begin{document}
\title{Phantom Forgetting: How BatchNorm State Fakes Deletion in Edge Split Machine Unlearning}

\author{
\IEEEauthorblockN{Anand Agrawal}
\IEEEauthorblockA{\textit{SUTD, Singapore}\\
Singapore\\
agrawal\_anand@sutd.edu.sg}
\and
\IEEEauthorblockN{Chillara Anil Kumar}
\IEEEauthorblockA{
\textit{SRM University-AP}, Andhra Pradesh\\India\\
anilkumar.c@srmap.edu.in}
\and
\IEEEauthorblockN{Mandan Naresh}
\IEEEauthorblockA{\textit{Manipal Institute of Technology, Manipal,}\\
India\\
mandan.naresh@manipal.edu}}

\maketitle

\begin{abstract}
Split learning is a natural fit for edge intelligence: a resource-constrained
edge device runs the first layers of a network and offloads the remainder to a
server, so that raw data never leave the device. When such a system must
\emph{unlearn} revoked or sensitive data for example to honor a
right-to-erasure request, the deletion is commonly assessed through changes in
forget-set accuracy. However, inference on an edge--server split also depends on
non-learnable model state distributed across the two parties, and this state is
especially volatile on edge devices subject to data drift, over-the-air updates,
and checkpoint rollback. This paper isolates a state-integrity question with
direct edge-security implications: can Batch Normalization (BN) running
statistics create a target-specific forgetting signal even when every learnable
weight remains fixed? We define a state-only diagnostic that separates
learnable-parameter change from BN-buffer change and localizes the effect to the
edge (client) side, the server side, and joint placements. We also introduce a
state-induced forgetting gap (SIFG) that discounts simultaneous retain-set
degradation. Over all ten CIFAR-10 forget classes, centralized BN recalibration
on forget data reduces mean forget accuracy by $61.50$ percentage points with
zero weight change; in a split model, edge-only, server-only, and joint
recalibration produce mean drops of $37.27$, $41.71$, and $51.90$ points. The
server-side state yields the largest selective gap relative to retain
degradation, while joint recalibration causes the largest overall prediction
change. These results establish distributed normalization buffers including
those held on the edge device as security and privacy-relevant inference
state, and show that forget accuracy should not be interpreted, or accepted as a
deletion certificate, without an explicit edge-aware state-integrity audit.
\end{abstract}


\section{Introduction}
Edge intelligence pushes model inference and, increasingly, model updates onto
resource-constrained devices such as phones, wearables, cameras, and IoT
sensors~\cite{shi2016edge,zhou2019edge,xu2020edge}. Because these devices cannot
host large models and must keep sensitive data local, a common design partitions
the network so that the device computes an early representation and a server
completes the remaining forward and backward computation. This partitioning
underlies both DNN offloading systems such as
Neurosurgeon~\cite{kang2017neurosurgeon} and split
learning~\cite{vepakomma2018,gupta2018,matsubara2022split}, where the client
sends a \emph{smashed activation} instead of raw input. The deployed prediction
therefore depends on state held on both sides of the edge--server communication
boundary.

Edge deployments must also support \emph{deletion}. Privacy regulation grants
individuals a right to erasure~\cite{gdpr2016}, and machine unlearning aims to
remove the influence of revoked or unauthorized training data without rebuilding
a model from scratch~\cite{cao2015,bourtoule2021,guo2020}. In distributed and
edge settings the problem is not only how to modify the model but also how to
\emph{verify} that an observed behavioral change represents removal rather than
an unrelated alteration of the deployed
system~\cite{sommer2022,thudi2022,gao2024verifi}. This is acute in split learning,
where recent work shows the systems cost of coordinating model-state updates
across the boundary~\cite{jiang2025} and where the edge client is often the least
trusted and least controlled component of the pipeline~\cite{xiao2019edgesec}.

A widely used verification signal is the reduction in accuracy on the forget
distribution: a large reduction is read as evidence that the target information
has been removed. This interpretation is incomplete for models that contain Batch
Normalization (BN). At inference time, BN uses running means and variances that
are updated from data but are not learnable parameters~\cite{ioffe2015};
re-estimating these buffers can alter downstream representations and predictions
without changing any trained weight. Test-time adaptation and post-training
recalibration confirm that normalization state can strongly influence behavior
under distribution shift~\cite{lim2023,tomar2024,su2024,shomron2020}, precisely
the regime edge devices operate in.

On an edge--server split the concern is amplified because normalization buffers
may reside on the edge client, on the server, or on both. A verifier that checks
only final outputs, or only weight tensors, may miss a behaviorally significant
change in the operational model. The risk needs no attacker: legitimate
recalibration, over-the-air (OTA) checkpoint updates applied non-atomically to a
device fleet, on-device data drift, or inconsistent client/server state can all
alter BN statistics. If the target-class accuracy then collapses, the change may
be misattributed to unlearning, an edge device may appear to have ``forgotten''
a user's data while the learned weights, and the information they encode, are
untouched.

This paper studies a deliberately narrow verification problem with concrete
edge-security consequences: whether distributed BN state can generate
target-specific apparent forgetting under zero learnable-weight change. The scope
is diagnostic rather than algorithmic. We do not propose a new unlearning
optimizer, compare complete unlearning mechanisms, or claim privacy-level
removal. Instead, we treat BN buffers as security and privacy-relevant inference state and test whether a
verifier that observes only forget accuracy can be misled by state updates.

The contributions are fourfold:
\begin{itemize}
\item We formulate an \emph{edge-aware state-integrity threat model} for split
machine unlearning in which learnable parameters remain fixed but edge-side or
server-side BN buffers drift.
\item We define a state-only diagnostic that reports parameter change and
BN-buffer change separately, together with forget, retain, test, and
prediction-disagreement metrics.
\item We introduce the \emph{state-induced forgetting gap} (SIFG), which measures
target-specific apparent forgetting after discounting retain-set degradation.
\item We localize the effect experimentally across centralized, edge-only,
server-only, and joint edge--server BN recalibration over all ten CIFAR-10 forget
classes, and translate the findings into edge-deployment failure scenarios and
minimum verification requirements.
\end{itemize}

\section{Related Work}\label{sec:rw}

\subsection{Edge Intelligence and Split Learning}
Edge computing moves computation close to data sources to reduce latency and
protect privacy~\cite{shi2016edge,zhou2019edge}. For deep models this motivates
partitioning the network between an edge device and a server: Neurosurgeon
selects a split point to balance device compute and
communication~\cite{kang2017neurosurgeon}, split computing surveys catalog such
device--server pipelines~\cite{matsubara2022split}, and split learning transmits
a smashed activation so that raw data stay on the
device~\cite{vepakomma2018,gupta2018}. SplitFed combines split and federated
learning for multi-device edge fleets~\cite{thapa2022splitfed}. Security studies
of the edge highlight that the on-device component and the device--server channel
form a distinct attack surface~\cite{xiao2019edgesec}. These works establish the
edge--server boundary as a first-class systems and security concern; we add that
non-learnable inference state on this boundary is itself a verification hazard.

\subsection{Machine-Unlearning Verification}
Early unlearning work focused on efficiently removing data influence or
approximating retraining without the deleted
samples~\cite{cao2015,bourtoule2021,guo2020}. Verification-oriented research later
emphasized that a behavioral change alone is not a sufficient definition of
removal. Athena evaluates probabilistic evidence of unlearning, and VeriFi
studies verifiable federated unlearning with observable evidence beyond a single
utility metric~\cite{sommer2022,gao2024verifi}; auditable definitions require the
evaluator to state what property is tested and which components may
change~\cite{thudi2022}. Split unlearning extends deletion to the partitioned
setting and quantifies the cost of coordinated cross-boundary
updates~\cite{jiang2025}. Our work follows this verification perspective but
focuses on a specific state confound, localized to the edge and server sides,
rather than proposing a general certification mechanism.

\subsection{Split-Learning Security and Normalization-State Sensitivity}
Security studies of split learning examine leakage and malicious behavior at the
device--server boundary, including pseudo-client functionality, data stealing,
and backdoor attacks against vertical split
learning~\cite{gao2023pcat,bai2023villain}. Our concern is complementary: even
without protocol compromise, non-learnable inference state can create a
misleading verification outcome. Separately, BN uses batch statistics during
training and running statistics during inference~\cite{ioffe2015}; these buffers
are stored in the model state but are not optimized by gradient descent.
Post-training BN recalibration can alter or recover performance, and test-time
normalization methods manage statistics under distribution
shift~\cite{lim2023,tomar2024,su2024,shomron2020}, conditions endemic to edge
fleets. We connect this behavioral sensitivity to split-unlearning verification
and localize it across the distributed edge--server boundary.

\section{Model State and Threat Model}\label{sec:model}

\subsection{Edge--Server Split Model}
Let the edge (client) model $f_c$ map an input $x$ to a smashed activation $z$,
and let the server model $f_s$ produce the output-score vector $s$:
\begin{equation}
z = f_c(x;\theta_c,\mathcal{B}_c),\qquad s = f_s(z;\theta_s,\mathcal{B}_s),
\end{equation}
where $\theta_c,\theta_s$ are learnable parameters and
$\mathcal{B}_c,\mathcal{B}_s$ are stateful normalization buffers held on the edge
device and the server, respectively. The complete model is
\begin{equation}
F(x;\theta,\mathcal{B}) = f_s\!\left(f_c(x;\theta_c,\mathcal{B}_c);\theta_s,\mathcal{B}_s\right),
\end{equation}
with $\theta=(\theta_c,\theta_s)$ and $\mathcal{B}=(\mathcal{B}_c,\mathcal{B}_s)$.
Here $F(x;\theta,\mathcal{B})\in\mathbb{R}^K$ is the score vector for $K$ classes
and $\hat{y}_F(x)=\arg\max_k [F(x;\theta,\mathcal{B})]_k$ is the predicted class.
Let $D_f$ be the forget set and $D_r$ the retain set. An unlearning-related
operation may modify learnable parameters, BN buffers, or both. A behavioral
verifier commonly observes overall test accuracy, retain accuracy, forget
accuracy, and prediction disagreement; these outputs do not identify whether a
change originates from $\theta$, $\mathcal{B}_c$, $\mathcal{B}_s$, or broad model
degradation.

\begin{table}[t]
\caption{Model-state cases relevant to edge--server verification.}
\label{tab:cases}
\centering
\footnotesize
\renewcommand{\arraystretch}{1.15}
\begin{tabular}{@{}ccll@{}}
\toprule
$\Delta_\theta$ & $\Delta_{\mathcal{B}}$ & Output & Interpretation\\
\midrule
$0$ & $0$ & None & Unchanged under fixed execution.\\
$>0$ & $0$ & Possible & Weight update with fixed BN.\\
$0$ & $>0$ & Possible & State-only apparent forgetting.\\
$>0$ & $>0$ & Possible & Mixed update; separate effects first.\\
\bottomrule
\end{tabular}
\end{table}

\subsection{BN Buffers as Inference State}
For an activation $a$, inference-time BN applies
\begin{equation}
\mathrm{BN}(a;\mu,\sigma^2,\gamma,\beta) =
\gamma\,\frac{a-\mu}{\sqrt{\sigma^2+\epsilon}} + \beta,
\label{eq:bn}
\end{equation}
where $\mu$ and $v=\sigma^2$ are channel-wise running-statistic buffers,
$\gamma,\beta$ are learnable affine parameters, and $\epsilon>0$ is a stability
constant. Let $(\theta',\mathcal{B}')$ denote the state after recalibration or
another model-state operation. We audit learnable state using
\begin{equation}
\Delta_\theta = \left(\lVert\theta'_c-\theta_c\rVert_2^2 + \lVert\theta'_s-\theta_s\rVert_2^2\right)^{1/2},
\end{equation}
where $\theta$ includes BN affine parameters and empty components are omitted for
a centralized model. For $j\in\{c,s\}$, let $\mathcal{L}^j_{\mathrm{BN}}$ be the
BN layers at location $j$ and define
\begin{equation}
\Delta_{\mu,j} = \!\!\!\sum_{\ell\in\mathcal{L}^j_{\mathrm{BN}}}\!\!\!\lVert\mu'_\ell-\mu_\ell\rVert_2,\quad
\Delta_{v,j} = \!\!\!\sum_{\ell\in\mathcal{L}^j_{\mathrm{BN}}}\!\!\!\lVert v'_\ell-v_\ell\rVert_2,
\end{equation}
with $\Delta_{\mathcal{B}_j}=\Delta_{\mu,j}+\Delta_{v,j}$ and
$\Delta_{\mathcal{B}}=\Delta_{\mathcal{B}_c}+\Delta_{\mathcal{B}_s}$. The
location-specific BN-state change used in the audit is
\begin{equation}
\Delta_{\mathcal{B}_L} =
\begin{cases}
\Delta_{\mathcal{B}_{\mathrm{all}}}, & L=\mathrm{all},\\
\Delta_{\mathcal{B}_c}, & L=c \ (\text{edge}),\\
\Delta_{\mathcal{B}_s}, & L=s \ (\text{server}),\\
\Delta_{\mathcal{B}_c}+\Delta_{\mathcal{B}_s}, & L=c{+}s.
\end{cases}
\end{equation}
For a centralized model $\Delta_{\mathcal{B}}=\Delta_{\mathcal{B}_{\mathrm{all}}}$.
The state-only diagnostic enforces $\Delta_\theta=0$ and
$\Delta_{\mathcal{B}}>0$. With data, preprocessing, non-BN state, and execution
settings fixed, any measured prediction change is attributable to the modified BN
running-state buffers.

\subsection{Edge-Aware Verification Threat}
\begin{figure}[t]
\centering
\resizebox{\columnwidth}{!}{%
\begin{tikzpicture}[
  font=\scriptsize,
  node distance=5mm,
  box/.style={draw, rounded corners=1pt, align=center, inner sep=3pt, minimum height=6mm},
  ebox/.style={box, fill=blue!7},
  sbox/.style={box, fill=orange!10},
  vbox/.style={box, fill=green!8},
  ar/.style={-latex, thick}
]
\node[ebox, minimum width=24mm] (edge) {Edge client\\$f_c(x;\theta_c,\mathcal{B}_c)$};
\node[sbox, minimum width=24mm, right=14mm of edge] (srv) {Server\\$f_s(z;\theta_s,\mathcal{B}_s)$};
\node[vbox, minimum width=24mm, right=10mm of srv] (pred) {Prediction\\ and metrics};
\draw[ar] (edge) -- node[above, font=\scriptsize] {$z$} (srv);
\draw[ar] (srv) -- (pred);
\node[box, fill=gray!8, minimum width=66mm, below=8mm of srv] (rec)
  {state-only recalibration: $\Delta_\theta\le\varepsilon_\theta$, $\Delta_{\mathcal{B}_L}>\varepsilon_{\mathcal{B}}$};
\draw[ar, dashed] (rec.north) -- (rec.north|-edge.south);
\draw[ar, dashed] (rec.north) -- (srv.south);
\node[box, fill=red!7, minimum width=66mm, below=5mm of rec] (q)
  {\textbf{Can BN state alone explain the forget drop?}\\
   audit $R_{\mathrm{state}}(L,q)$: behavior, $\Delta_\theta$, location-specific BN drift};
\draw[ar] (rec) -- (q);
\end{tikzpicture}}
\caption{Edge-aware state-integrity diagnostic. Selected edge-client and/or
server BN buffers are recalibrated while all learnable parameters remain frozen,
and the audit asks whether distributed BN state alone can account for an observed
forget-accuracy drop.}
\label{fig:diag}
\end{figure}
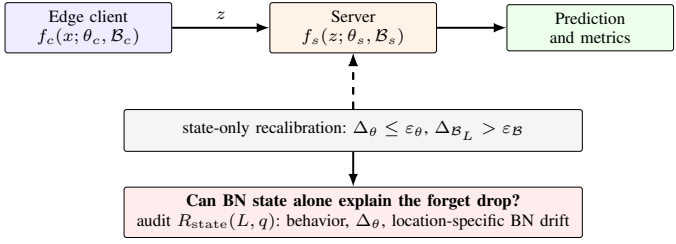

The verifier may be a system operator, internal auditor, compliance authority, or
external evaluator. We assume no poisoning, model extraction, test manipulation,
or protocol compromise. The threat arises when a legitimate state transition of
the kind routinely produced by edge operation is interpreted incorrectly.
Table~\ref{tab:cases} distinguishes four cases. A weak verifier accepts a large
forget-accuracy reduction as successful removal. A state-aware verifier instead
asks whether the same behavior occurs in the third row of Table~\ref{tab:cases}:
if it does, forget accuracy is confounded by normalization state and cannot
independently identify parameter-level unlearning. Because the edge client is
frequently updated out of band and rarely re-audited weight-by-weight, its BN
buffers are a particularly plausible and particularly consequential source of
this confound. Fig.~\ref{fig:diag} summarizes the diagnostic.

\section{State-Integrity Diagnostic}\label{sec:diag}

\subsection{Behavioral Reductions and the SIFG}
For a model state $G$, let $\hat{y}_G(x)=\arg\max_k[G(x)]_k$ and define
\begin{align}
A(G;S) &= \frac{1}{|S|}\!\!\sum_{(x,y)\in S}\!\!\mathbb{I}[\hat{y}_G(x)=y],\\
D_{\mathrm{pred}}(F,G) &= \frac{1}{|D^{\mathrm{test}}|}\!\!\sum_{(x,y)\in D^{\mathrm{test}}}\!\!\mathbb{I}[\hat{y}_F(x)\neq\hat{y}_G(x)].
\end{align}
We use $A_f(G)=A(G;D^{\mathrm{test}}_f)$, $A_r(G)=A(G;D^{\mathrm{test}}_r)$, and
$A_{\mathrm{test}}(G)=A(G;D^{\mathrm{test}})$. Let $F_0$ be the corresponding
unrecalibrated reference model and let $F^{\mathrm{BN}}_{L,q}$ denote the state
obtained by recalibrating location $L\in\{\mathrm{all},c,s,c{+}s\}$ using
$D^{\mathrm{cal}}_q$, $q\in\{f,r\}$. For $u\in\{f,r\}$,
\begin{equation}
\delta^{\mathrm{BN}}_u(L,q) = A_u(F_0) - A_u(F^{\mathrm{BN}}_{L,q}).
\end{equation}
A large $\delta^{\mathrm{BN}}_f(L,q)$ under zero parameter change demonstrates
that a behavioral forgetting signal can be generated without modifying learned
weights. To discount retain degradation we define the \emph{state-induced
forgetting gap}
\begin{equation}
\mathrm{SIFG}(L,q) = \delta^{\mathrm{BN}}_f(L,q) - \delta^{\mathrm{BN}}_r(L,q).
\label{eq:sifg}
\end{equation}
A positive SIFG means the forget distribution degrades more strongly than the
retain distribution. For a pre-specified threshold $\tau_f>0$ and tolerances
$\varepsilon_\theta,\varepsilon_{\mathcal{B}}\ge0$, an observation is
\emph{state-confounded} when
\begin{equation}
\begin{aligned}
&\delta^{\mathrm{BN}}_f(L,q)\ge\tau_f,\qquad \mathrm{SIFG}(L,q)>0,\\
&\Delta_\theta\le\varepsilon_\theta,\qquad \Delta_{\mathcal{B}_L}>\varepsilon_{\mathcal{B}}.
\end{aligned}
\label{eq:conf}
\end{equation}
In the controlled experiments parameter equality is exact, so
$\varepsilon_\theta=0$. This label is diagnostic rather than a privacy or removal
certificate.

\subsection{Localization States}
We evaluate centralized ($L=\mathrm{all}$), edge-only ($L=c$), server-only
($L=s$), and joint edge--server ($L=c{+}s$) recalibration. Non-selected BN layers
preserve their original buffers, and a retain-data control is included centrally
to distinguish calibration-distribution effects from repeated evaluation. The
location-specific report is
\begin{equation}
\begin{aligned}
R_{\mathrm{state}}(L,q) = \big(&A_f,\, A_r,\, A_{\mathrm{test}},\, D_{\mathrm{pred}}(F_0,F^{\mathrm{BN}}_{L,q}),\\
&\Delta_\theta,\, \Delta_{\mathcal{B}_c},\, \Delta_{\mathcal{B}_s},\, \mathrm{SIFG}\big),
\end{aligned}
\label{eq:report}
\end{equation}
evaluated at $F^{\mathrm{BN}}_{L,q}$; for $L=\mathrm{all}$ the pair
$(\Delta_{\mathcal{B}_c},\Delta_{\mathcal{B}_s})$ collapses to the single quantity
$\Delta_{\mathcal{B}_{\mathrm{all}}}$. This optimizer-independent report tests
whether deployed state provides an alternative explanation for the observed
behavior.

\begin{algorithm}[t]
\caption{Location-Aware BN State Audit}
\label{alg:audit}
\begin{algorithmic}[1]
\REQUIRE $F_0$, $D^{\mathrm{cal}}_q$ ($q\in\{f,r\}$), $L\in\{\mathrm{all},c,s,c{+}s\}$
\STATE Copy $F_0$ to $F^{\mathrm{BN}}_{L,q}$; freeze all learnable tensors.
\STATE Set the model to evaluation mode; set only BN modules at $L$ to training mode.
\STATE Traverse $D^{\mathrm{cal}}_q$ under no-gradient execution to update selected running statistics.
\STATE Restore evaluation mode and compute $R_{\mathrm{state}}(L,q)$.
\ENSURE $\Delta_\theta\le\varepsilon_\theta$ and every non-selected buffer is unchanged.
\end{algorithmic}
\end{algorithm}

\section{Experimental Method}\label{sec:method}

\subsection{Dataset and Forget Distributions}
We use CIFAR-10~\cite{krizhevsky2009}. For each class $c_f\in\{0,\dots,9\}$,
training samples with label $c_f$ form the forget distribution and the remaining
classes form the retain distribution. Let $D^{\mathrm{cal}}_q\subseteq
D^{\mathrm{train}}_q$, $q\in\{f,r\}$, be the calibration samples used to update BN
running statistics; held-out test samples are used only for evaluation and never
for recalibration. Forget and retain accuracies are measured on their
corresponding held-out test partitions. Reported values are mean $\pm$ standard
deviation across the ten choices of $c_f$; the variation therefore represents
class-wise sensitivity rather than random-seed variation.

\subsection{Models and Recalibration}
We reuse fixed 50-epoch checkpoints and a common data pipeline; no checkpoint is
retrained or tuned for this study. The centralized diagnostic uses a compact BN
convolutional network. In the split network the edge client contains 32- and
64-channel BN feature blocks and transmits the 64-channel smashed activation,
while the server contains a 128-channel BN block followed by classification. This
placement permits independent edge-only, server-only, and joint recalibration and
mirrors an edge deployment in which shallow layers run on-device.

For every state-only experiment, all learnable tensors, including BN affine
parameters, are frozen. The complete model is first placed in evaluation mode,
after which only the selected BN modules are placed in training mode. A single
gradient-free traversal of the designated training partition updates the selected
running means and variances: $|D^{\mathrm{cal}}_f|=5000$ for the current forget
class and $|D^{\mathrm{cal}}_r|=45000$ for the centralized retain control. Each
derived state starts from its corresponding unchanged checkpoint $F_0$. Full
model-state dictionaries are retained for integrity checking; the reported
$\Delta_{\mathcal{B}}$ includes the inference-active running-mean and
running-variance buffers, while framework bookkeeping buffers are checked
separately and excluded from the drift score.

\subsection{Audit Procedure}
Algorithm~\ref{alg:audit} derives every recalibrated state from its corresponding
$F_0$ and \emph{verifies} parameter equality rather than assuming that disabling
the optimizer is sufficient. With evaluation data, preprocessing, non-BN state,
and execution settings fixed, $\Delta_\theta=0$ and unchanged non-selected
buffers attribute the measured difference to the selected BN location. We report
test, retain, and forget accuracy because a target-class collapse may coexist
with broad utility loss, and we additionally report prediction disagreement
because accuracy can hide substantial label changes that cancel in aggregate. The
SIFG derived from forget and retain reductions captures selectivity but is not
used to suppress the underlying metrics. The exact scripts, configuration, and
per-class result summaries are maintained in a version-controlled reproducibility
package for release with the final paper.

\section{Results}\label{sec:results}
Table~\ref{tab:main} summarizes the centralized and split-location experiments.

\begin{table*}[t]
\caption{State-only BN recalibration over ten CIFAR-10 forget classes. Accuracies and disagreement are mean $\pm$ standard deviation; reductions and SIFG are differences of the reported means. Split rows use forget-data recalibration, and all recalibrated states satisfy $\Delta_\theta=0$.}
\label{tab:main}
\centering
\footnotesize
\renewcommand{\arraystretch}{1.1}
\begin{tabular}{@{}llccccccc@{}}
\toprule
Setting & State & Test & Retain & Forget & $\delta^{\mathrm{BN}}_f$ & $\delta^{\mathrm{BN}}_r$ & SIFG & $D_{\mathrm{pred}}$\\
\midrule
\multirow{3}{*}{Centralized}
 & Original & $0.7425{\pm}0.0152$ & $0.7433{\pm}0.0193$ & $0.7357{\pm}0.1016$ & -- & -- & -- & $0.0000{\pm}0.0000$\\
 & Forget recal. & $0.6036{\pm}0.0346$ & $0.6573{\pm}0.0384$ & $0.1207{\pm}0.0289$ & $0.6150$ & $0.0860$ & $\mathbf{0.5290}$ & $0.3361{\pm}0.0438$\\
 & Retain recal. & $0.7445{\pm}0.0109$ & $0.7373{\pm}0.0150$ & $0.8096{\pm}0.0755$ & $-0.0739$ & $0.0060$ & $-0.0799$ & $0.1791{\pm}0.0211$\\
\midrule
\multirow{4}{*}{Split (edge/server)}
 & Original & $0.7427{\pm}0.0114$ & $0.7428{\pm}0.0186$ & $0.7411{\pm}0.1170$ & -- & -- & -- & $0.0000{\pm}0.0000$\\
 & Edge BN ($c$) & $0.6061{\pm}0.0392$ & $0.6325{\pm}0.0434$ & $0.3684{\pm}0.2225$ & $0.3727$ & $0.1103$ & $0.2624$ & $0.3468{\pm}0.0507$\\
 & Server BN ($s$) & $0.6821{\pm}0.0446$ & $0.7219{\pm}0.0481$ & $0.3240{\pm}0.0880$ & $0.4171$ & $0.0209$ & $\mathbf{0.3962}$ & $0.2505{\pm}0.0634$\\
 & Joint BN ($c{+}s$) & $0.5575{\pm}0.0713$ & $0.5948{\pm}0.0717$ & $0.2221{\pm}0.1538$ & $\mathbf{0.5190}$ & $0.1480$ & $0.3710$ & $\mathbf{0.4195{\pm}0.0948}$\\
\bottomrule
\end{tabular}
\end{table*}

\subsection{Centralized State-Only Apparent Forgetting}
The original centralized model attains mean forget accuracy $0.7357$.
Recalibrating only BN running statistics on forget data reduces it to $0.1207$, a
$61.50$-point drop, while $\Delta_\theta=0$. Retain accuracy decreases by only
$8.60$ points, giving a SIFG of $0.5290$, and the post-recalibration model changes
its prediction on $33.61\%$ of test samples. The retain-data control behaves
differently: recalibration on retain samples leaves test and retain accuracy close
to the original values and \emph{increases} mean forget accuracy to $0.8096$. Its
negative SIFG shows that the calibration distribution, not repeated evaluation
alone, determines the effect. For any $\tau_f\le0.6150$ and
$\varepsilon_{\mathcal{B}}$ below the measured BN change, the centralized
forget-data result satisfies~\eqref{eq:conf} with exact parameter equality
($\varepsilon_\theta=0$).

\subsection{Edge and Server Localization}
In the split model both sides independently produce substantial forget-class
collapse. Edge-only recalibration reduces forget accuracy from $0.7411$ to
$0.3684$, and server-only recalibration reduces it to $0.3240$; the corresponding
forget drops are $0.3727$ and $0.4171$. The server-side state produces the
largest selective effect: retain accuracy decreases by only $0.0209$, giving a
SIFG of $0.3962$. Thus server BN buffers can create a strong target-specific
behavioral signal even though they are downstream of the smashed-activation
boundary. The edge-only setting produces more prediction disagreement but also
more retain degradation, yielding a smaller SIFG of $0.2624$. Joint recalibration
produces the largest overall forget drop, $0.5190$, and the largest prediction
disagreement, $0.4195$; its SIFG, $0.3710$, is lower than the server-only value
because retain accuracy also falls by $0.1480$. The placement causing the largest
accuracy collapse is therefore not necessarily the placement causing the most
selective state-only effect---an important distinction when an edge fleet and a
server can drift independently.

\begin{table}[t]
\caption{Centralized BN-state drift and behavior (ten-class means). $\Delta_{\mathcal{B}}=\Delta_\mu+\Delta_v$.}
\label{tab:drift}
\centering
\footnotesize
\renewcommand{\arraystretch}{1.15}
\begin{tabular}{@{}lccccc@{}}
\toprule
Calibration & $\Delta_\mu$ & $\Delta_v$ & $\Delta_{\mathcal{B}}$ & $\delta^{\mathrm{BN}}_f$ & $D_{\mathrm{pred}}$\\
\midrule
Forget & $1.5388$ & $3.7875$ & $5.3263$ & $0.6150$ & $0.3361$\\
Retain & $1.2092$ & $3.6197$ & $4.8289$ & $-0.0739$ & $0.1791$\\
\bottomrule
\end{tabular}
\end{table}

\subsection{Class-Wise Stability and Disagreement}
Paired class-wise forget-drop standard deviations, computed directly from the ten
per-class reductions, are $0.1299$ (server-only), $0.2633$ (edge-only), and
$0.2077$ (joint). Server-side apparent forgetting is therefore not only selective
on average but also more stable across classes in this model, whereas edge-side
sensitivity varies substantially with the selected class. The disagreement
statistics reinforce this: server-only recalibration changes fewer test
predictions than edge-only ($0.2505$ vs.\ $0.3468$) while producing a
\emph{larger} forget drop, consistent with the SIFG reading that the server update
is more concentrated on the target behavior. Joint recalibration changes the
largest fraction of predictions, indicating that simultaneous state movement
across the edge--server boundary has the broadest effect.

\subsection{State-Change Magnitude Does Not Predict Behavior}
Table~\ref{tab:drift} reports the centralized running-mean and running-variance
drift. Forget-data recalibration has magnitude $5.3263$, only about $10.3\%$ above
the retain-data value $4.8289$, yet their forget-accuracy effects are
\emph{opposite}: $0.6150$ versus $-0.0739$. The drift score $\Delta_{\mathcal{B}}$
confirms that state changed but does not reveal which activation distribution
shaped the statistics or which decisions will be affected. Similar magnitudes can
point in different directions in state space and produce different class behavior,
so state-change thresholds must be accompanied by calibration provenance and
behavioral measurements---an edge integrity check that hashes buffer sizes or
norms alone is insufficient.

\subsection{Verification Consequences}
The results establish four points. Forget accuracy is \emph{state-confounded}:
drops above $50$ points occur with unchanged learnable parameters. The confound is
\emph{distributed}: either edge or server buffers can independently change
predictions. Calibration \emph{provenance matters}: forget-data and retain-data
recalibration have opposite target effects. Finally, collapse magnitude and
selectivity differ: joint recalibration maximizes the forget drop, whereas
server-only recalibration maximizes SIFG. These findings motivate a minimum
state-integrity report for split machine unlearning. A large forget drop
accompanied by $\Delta_\theta=0$ should be labeled state-induced apparent
forgetting, not certified removal, and the report is compatible with any
unlearning optimizer because it evaluates the deployed model state rather than
prescribing how the model must be updated.

\section{Edge Operational Failure Scenarios}\label{sec:scenarios}
The state-only mechanism can appear in several edge deployment scenarios even when
the intended unlearning algorithm does not modify BN buffers.

\textbf{Evaluation-mode leakage on-device.} An edge client or server component may
remain in training mode during post-update validation. Forwarding
forget-distribution batches then changes running statistics even though gradients
and optimizer steps are disabled. The evaluator may believe the model is frozen
because $\theta$ is unchanged, while $\mathcal{B}_c$ or $\mathcal{B}_s$ continues
to evolve. Algorithm~\ref{alg:audit} treats this as an explicit counterfactual; a
production pipeline should prevent it unless recalibration is deliberately
requested.

\textbf{Non-atomic OTA deployment across a device fleet.} Split models are shipped
as multiple artifacts, and OTA updates to a heterogeneous edge fleet rarely land
atomically. An edge checkpoint may be rolled back while the server retains newer
BN statistics, or the server may restore weights without restoring its buffers.
The resulting pair does not correspond to any coherently trained model state.
Because edge-only and server-only recalibration independently change forget
behavior, a non-atomic deployment can create or suppress an apparent unlearning
effect; a common version identifier must bind both sides and all state components.

\textbf{Incomplete on-device integrity measurement.} A device-attestation or
model-integrity mechanism may hash serialized parameter tensors but exclude
registered buffers, reporting the same identity for states that share $\theta$ but
differ in $\mathcal{B}$. Our experiments show that such states can disagree on more
than $40\%$ of test predictions under joint recalibration. Integrity metadata for
an unlearning audit must therefore cover running means, running variances, and
other persistent inference buffers, not only trainable tensors and must do so on
the edge device, where measurement is often weakest.

\textbf{Deliberate threshold manipulation.} Although our primary threat model
requires no adversary, the same mechanism can be exploited. An operator seeking to
satisfy a forget-accuracy threshold could recalibrate BN on target-distribution
samples while preserving all learned weights; a verifier that checks only $A_f$
would observe a strong forgetting signal. Verifying $\Delta_\theta$, buffer
provenance, retain behavior, and location-specific counterfactuals prevents this
state-only change from being presented as parameter-level removal.

\section{State-Aware Verification Requirements}\label{sec:req}
The measurements above convert into four concrete requirements for a split- and
edge-unlearning verifier, summarized in Table~\ref{tab:req}. A model identifier
should bind $\theta$, $\mathcal{B}_c$, and $\mathcal{B}_s$ and confirm that edge
and server snapshots belong to one coherent version. Location-aware
counterfactuals should restore or recalibrate BN state one side at a time, because
either component can independently generate apparent forgetting. Audit records
should also identify the calibration source, updated layers, sample count, and
batch ordering. The acceptance decision must combine these state records with
retain utility and prediction disagreement. When~\eqref{eq:conf} holds, the
supported conclusion is BN-confounded behavior; actual removal requires additional
method-specific evidence.

\begin{table}[t]
\caption{Minimum state-aware verification requirements for edge--server split unlearning.}
\label{tab:req}
\centering
\footnotesize
\renewcommand{\arraystretch}{1.2}
\begin{tabular}{@{}p{16mm}p{28mm}p{28mm}@{}}
\toprule
Requirement & Evidence & Prevented failure\\
\midrule
State identity & Version $\theta$, $\mathcal{B}_c$, $\mathcal{B}_s$ together & Mixed or stale edge/server checkpoint accepted as coherent\\
Location control & Edge, server, and joint diagnostics & Change attributed to the wrong partition\\
Provenance & Record source, count, order, and layers & Hidden or accidental buffer update\\
Joint decision & Report Eq.~\eqref{eq:report}, not $A_f$ alone & State-induced collapse certified as removal\\
\bottomrule
\end{tabular}
\end{table}

\section{Security Discussion and Limitations}\label{sec:disc}

\subsection{Edge State Integrity as a Trust Boundary}
Split-learning security work commonly emphasizes raw-data confidentiality,
smashed-activation leakage, and malicious
protocols~\cite{gao2023pcat,bai2023villain,xiao2019edgesec}. Our results identify
persistent inference state held partly on the edge device as a complementary
trust boundary. Authenticating only parameter tensors misses deployments that
share $\theta$ but differ in BN buffers and behavior. The issue needs no
adversary: on-device recalibration, distribution shift, recovery, or inconsistent
checkpoint restoration across a fleet can produce false attribution. A practical
edge verifier should preserve a pre-update snapshot of $\theta$, $\mathcal{B}_c$,
and $\mathcal{B}_s$, compare them separately, join edge/server reports by one
version identifier, and execute Algorithm~\ref{alg:audit} when possible. Where the
edge device supports a trusted execution environment, binding the BN-buffer digest
into device attestation would close the gap that parameter-only integrity checks
leave open.

\subsection{Scope and Limitations}
The study isolates mechanism and location; it does not compare retraining or
gradient-based unlearning, evaluate non-IID edge clients, or perform
membership-inference audits. It uses CIFAR-10, compact convolutional models, and
class-wise rather than multi-seed aggregation. Additional datasets, architectures,
normalization layers, seeds, heterogeneous edge fleets, and device-level integrity
and attestation tests are needed to establish generality. SIFG is a behavioral
diagnostic, not a privacy certificate, and must be combined with
algorithm-specific and privacy evidence for data-removal claims.

\section{Conclusion}\label{sec:concl}
BatchNorm running statistics are distributed, security- and privacy-relevant state
in split machine unlearning, and part of that state lives on the edge device. With
learnable parameters frozen, forget-data recalibration caused a $61.50$-point
centralized drop and edge-only, server-only, and joint drops of $37.27$, $41.71$,
and $51.90$ points; server-side BN state had the highest SIFG. Forget accuracy
alone cannot distinguish learned-weight removal from state drift, so
edge--server split-unlearning evaluation must audit both the edge-client and the
server BN buffers, bind them under a single version identity, and treat a
forget-accuracy collapse under $\Delta_\theta=0$ as state-induced apparent
forgetting rather than certified deletion.

\end{document}